\documentclass[aps,prl,twocolumn,amsmath,amssymb,floatfix,nofootinbib,groupedaddress]{revtex4-1}

\newcommand{\nhat}{\hat{ \mathbf{n}}}

\usepackage{graphicx}
\usepackage{dcolumn}
\usepackage{bm}
\usepackage[usenames, dvipsnames]{color}
\usepackage[normalem]{ulem}

\graphicspath{ {figures/} }

\newcommand{\be}{\begin{eqnarray}}

\newcommand{\ee}{\end{eqnarray}}

\begin{document}

\title{Reconstructing the Primary CMB Dipole}
\newcommand{\cita}{Canadian Institute for Theoretical Astrophysics, University of Toronto, 60 St.~George Street, Toronto, Canada}

\author{P. Daniel Meerburg}

\author{Joel Meyers}

\author{Alexander van Engelen}
\affiliation{\cita}
\date{\today}

\begin{abstract}
The observed dipole anisotropy of the cosmic microwave background (CMB) temperature is much larger than the fluctuations observed on smaller scales and is dominated by the kinematic contribution from the Doppler shifting of the monopole due to our motion with respect to the CMB rest frame.  In addition to this kinematic component, there is expected to be an intrinsic contribution with an amplitude about two orders of magnitude smaller.  Here we explore a method whereby the intrinsic CMB dipole can be reconstructed through observation of temperature fluctuations on small scales which result from gravitational lensing.  Though the experimental requirements pose practical challenges, we show that one can in principle achieve a cosmic variance limited measurement of the primary dipole using the reconstruction method we describe.  Since the primary CMB dipole is sensitive to the largest observable scales, such a measurement would have a number of interesting applications for early universe physics, including testing large-scale anomalies, extending the lever-arm for measuring local non-Gaussianity, and constraining isocurvature fluctuations on super-horizon scales. 

\end{abstract}

\maketitle
\section{Introduction}
Measurements of the dipole anisotropy of the cosmic microwave background (CMB) temperature have a long history~\cite{Lineweaver:1996qw}. The first measurement of the dipole amplitude was made from the ground in 1969~\cite{1969Natur.222..971C}, and the first measurement of the direction and amplitude was made from a U2 jet aircraft in 1977~\cite{Smoot:1977bs}.  Observations from satellites have greatly improved the precision with which we measure the CMB dipole anisotropy~\cite{Kogut:1993ag,Fixsen:1996nj,Hinshaw:2008kr,Adam:2015vua}. 
The most precise measurement to date was obtained by the $\textit{Planck}$ satellite, resulting in a constraint of $\Delta T_1^{\mathrm{obs}} = 3364.3 \pm 1.5~\mu$K \cite{Adam:2015vua}. In general, the CMB dipole contains both a kinematic component, resulting from the Doppler boosting of the monopole, and an intrinsic component, due to the presence of large scale primordial fluctuations. 

It is typically assumed that the kinematic component is responsible for about 99\% of the observed dipole amplitude, which when combined with measurements of the CMB monopole~\cite{Fixsen:1996nj,Fixsen:2009ug} implies that our solar system has a velocity of roughly 370~km/s with respect to the CMB rest frame.

In addition to generating the kinematic dipole, the Doppler boosting due to our motion results in both frequency-independent aberration and frequency-dependent modulation of observed CMB fluctuations~\cite{Challinor:2002zh}.  The former has the same observational effect as a dipolar gravitational lensing potential while the latter has an effect similar to a dipole in the anisotropic optical depth, however with a differing frequency-dependence.  These effects of statistical anisotropy result in mode-coupling of anisotropies on different angular scales for a full-sky survey. While aberration results purely from our motion relative to the CMB, modulation occurs in the presence of any large scale temperature anisotropy and so is not uniquely sensitive to the kinematic dipole~\cite{Aghanim:2013suk,Roldan:2016ayx}.
The mode-coupling which results can be used to reconstruct our velocity relative to the local CMB rest frame and thereby the kinematic dipole~\cite{Amendola:2010ty,Abberation2011}.
An analysis of aberration and modulation with CMB maps from the \textit{Planck} satellite yielded an amplitude and direction for our motion which was consistent with that obtained from  direct measurement of the CMB dipole, assuming it is purely kinematic~\cite{Aghanim:2013suk}. 

One is always free to choose a set of coordinates in which the CMB dipole precisely vanishes. 
This frame does not in general coincide with the CMB rest frame, and will not have vanishing aberration of CMB fluctuations~\cite{Lewis:2006fu,Zibin:2008fe}.  We define the local CMB rest frame to be the one in which the locally observed aberration effect vanishes; the dipole observed in this frame is what we refer to as the primary or intrinsic CMB dipole.  The observed amplitude of fluctuations on smaller scales suggests that the intrinsic CMB dipole should have an amplitude of roughly 
$\Delta T_1^{\mathrm{prim}} \sim 30~ \mu$K for a standard cosmology.  
Constraints on the primary dipole could be derived by subtracting from the observed CMB dipole the kinematic dipole, the latter of which can be inferred from the aberration of the CMB~\cite{Aghanim:2013suk,Challinor:2002zh,Amendola:2010ty}, measurements of the clustering dipole from galaxy surveys~\cite{2010PhRvD..82d3530I,Yoon:2015lta},  or the frequency dependence of the motion-induced quadrupole ~\cite{Kamionkowski:2002nd,Yasini:2016dnd,Balashev:2015lla}.  It is also possible to gain insight using observations of the kinetic Sunyaev-Zel'dovich (kSZ) effect for distant clusters~\cite{Yasini:2016pby,Terrana:2016xvc}.

An observation of the primary CMB dipole would provide an independent measurement of the largest physical scales and would thus yield a number of applications to early universe physics.  Several large scale CMB anomalies have been identified in the data~\cite{Schwarz:2004gk,Gruppuso:2013dba,Ade:2015hxq,Schwarz:2015cma}, and cosmological models aimed at explaining these features would greatly benefit from the additional data the intrinsic dipole would provide.  
Super-horizon adiabatic fluctuations do not contribute to the primordial dipole, up to corrections which are suppressed by $(k\chi_\star)^2$, with  $k$ the comoving wavenumber and $\chi_\star$ the comoving distance to the CMB last-scattering surface~\cite{Erickcek:2008jp,Zibin:2008fe}.
Large-scale isocurvature fluctuations do not experience the same suppression, and so constraints on the primordial dipole would place constraints on such isocurvature modes.  Finally, models which contain local primordial non-Gaussianity predict a coupling between large- and small-scale fluctuations, and observational constraints on local non-Gaussianity would improve by extending to the largest possible scales.

In this paper, we propose a technique to reconstruct the primary dipole using the properties of gravitational lensing.  Lensing of the CMB by potential fluctuations along the line of sight generates small scale fluctuations in the presence of large scale gradients.  Upcoming CMB surveys will provide high-fidelity maps of both the small-scale CMB temperature field and the CMB lensing field \cite{Henderson:2015nzj,Benson:2014qhw,Suzuki:2015zzg,CMBS4, 2015hsa8.conf..334R}.
Our lensing-based method follows a similar procedure recently proposed to reconstruct large-scale polarization derived in Ref.~\cite{Meerburg:2017lfh}.  Below, we compute forecasts for reconstructing the primary dipole, and in an appendix we argue that this technique is insensitive to the kinematic dipole.  We also compare with the alternative aberration-based method for the primary dipole, which is obtained by differencing the direct dipole measurement from a velocity reconstruction based on the kinematic aberration effect.

\begin{figure}
\includegraphics[width=\columnwidth]{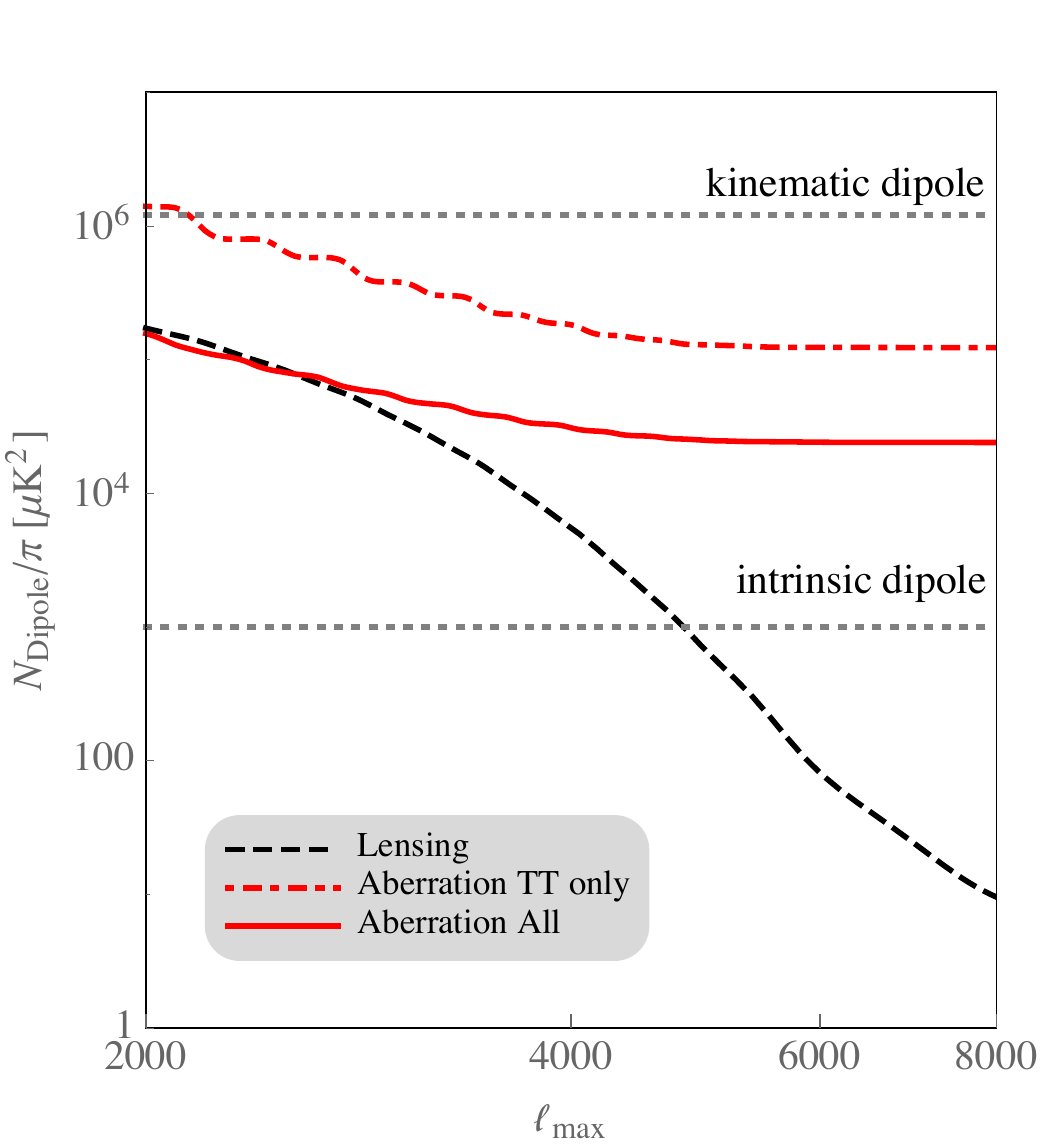}
\caption{Noise on the intrinsic dipole (Eq.~\ref{eq:reconstructionnoise}) for cosmic variance-limited CMB and lensing surveys up to a given $\ell_{\rm max}$, assuming delensing has been performed perfectly. The black dashed line shows the noise using our lensing-based method, for which a signal to noise ratio of 1 is attainable with $\ell_{\rm max} \simeq 5000$. In red, we show the noise attainable using the aberration-based method by differencing the observed dipole and the kinematic reconstruction, using only the $TT$ reconstruction (dashed) and using all quadratic combinations, i.e. $TT$, $TE$, $EE$ and $BB$ (solid). The dotted curves show  the amplitude of the kinematic dipole and the expected primary dipole in $\Lambda$CDM.}
\label{fig:PerfectUniverseDipole}
\end{figure}

\section{Reconstruction}
Gravitational lensing by matter between us and the last scattering surface distorts the temperature field in the direction $\nhat$ according to $\widetilde{T}(\nhat) = T(\nhat + \nabla \phi(\nhat))$ with $T$ the temperature field and $\phi$ the gravitational lensing potential. 
This distortion in temperature can be written as   \cite{Okamoto:2003zw} 
\be 
	\widetilde{T}_{\ell_1 m_1}^* = \sum_{\ell_2 m_2 \ell_3 m_3} \Gamma^{\ell_1 \ell_2 \ell_3}_{m_1 m_2 m_3} \phi_{\ell_2 m_2} T_{\ell_3 m_3} \, ,\label{eq:tonereconstruction}
\ee
where
\be 
    \Gamma^{\ell_1 \ell_2 \ell_3}_{m_1 m_2 m_3} = (-i) e_{\ell_1 \ell_2 \ell_3} I^{ \ell_1 \ell_2 \ell_3}_{m_1 m_2 m_3}\, ,
\ee
and $e_{\ell_1 \ell_2 \ell_3}$ is equal to unity when the sum $\ell_1 + \ell_2 + \ell_3$ is even and zero when the sum is odd.  
The mode-coupling integral $I$ is given by
\be 
	I^{\ell_1 \ell_2 \ell_3}_{m_1 m_2 m_3} &=& \sqrt{\frac{(2\ell_1 + 1)(2\ell_2+1)(2\ell_3+1)}{4\pi}} \nonumber \\ 
    & &\times  J_{\ell_1 \ell_2 \ell_3} \left( \begin{matrix} \ell_1 & \ell_2 & \ell_3 \\ m_1 & m_2 & m_3 \end{matrix} \right) \,, 
\ee
with
\be
	J_{\ell_1 \ell_2 \ell_3} &=&   \frac{-\ell_1(\ell_1+1)+\ell_2(\ell_2+1) + \ell_3(\ell_3+1)}{2} \nonumber \\ 
	& & \times \left( \begin{matrix} \ell_1 & \ell_2 & \ell_3 \\ 0 & 0 & 0 \end{matrix} \right) .
\ee
As described in the appendix, the kinematic dipole is not lensed by stationary lenses, and so the small scale lensed temperature is sensitive only to the intrinsic dipole.

Our estimate for the temperature dipole  derived from the observed, lensed temperature map $T^{\mathrm{obs}}$ and the observed lensing field $\phi^{\mathrm{obs}}$ is  given by
\be 
	\widehat{T}_{1 m} = \sum_{\ell_1 m_1 \ell_2 m_2} W^{ \ell_1 \ell_2 1 }_{m_1 m_2 m} T^{\mathrm{obs}*}_{\ell_1 m_1} \phi^{\mathrm{obs}*}_{\ell_2 m_2} \, 
\ee
with weights $W$ that we will determine below.  
Our lensing-based approach can be written schematically as $\widehat T_{1m} \sim \langle T_{\ell m^\prime} \phi_{(\ell \pm 1), m^{\prime\prime}}\rangle$; this can be contrasted with the reconstruction of our velocity using mode-couplings from the aberration and modulation effects, which are both obtained as  $\widehat \beta_{1m} \sim \langle T_{\ell m^\prime} T_{(\ell \pm 1), m^{\prime\prime}}\rangle$.

We  choose the weights $W$ to minimize the variance of the reconstruction, subject to the constraint that our estimate be unbiased $\left\langle\widehat{T}_{1 m} \right \rangle = T_{1 m}$,
\be 
	\sum_{\ell_1 m_1 \ell_2 m_2} W^{ \ell_1 \ell_2 1}_{m_1 m_2 m} \Gamma^{ \ell_1 \ell_2 1}_{m_1 m_2 m} C_\ell^{\phi\phi} = 1 \, . 
\ee
The variance of this estimator can then be written as \cite{Meerburg:2017lfh} 
\be 
	&\mathrm{Var}&\left(\widehat{T}_{1 m}\right) = \sum_{\ell_1 m_1 \ell_2 m_2} W^{\ell_1 \ell_2 1}_{m_1 m_2 m} W^{\ell_1 \ell_2 1 *}_{m_1 m_2 m}  \nonumber \\
    & \times & \left(C_{\ell_1}^{TT}+N_{\ell_1}^{TT}\right) \left(C_{\ell_2}^{\phi\phi}+N_{\ell_2}^{\phi\phi}\right) \, ,
\ee
where $N_{\ell}^{TT}$ is the noise on the CMB temperature and $N_{\ell}^{\phi\phi}$ is the noise on the lensing map, the latter of which can be obtained either internally from the CMB \cite{Hu:2001kj,Okamoto:2003zw} or by using a large-scale structure tracer as a proxy \cite{Smith:2010gu,Simard:2014aqa,Sherwin:2015baa}.
We compute the weights which minimize the noise in the reconstruction, yielding 
\be 
	W^{ \ell_1 \ell_2 1}_{m_1 m_2 m} &=& N_{\rm Dipole}^{\widehat{T} \widehat{T}}\left(\frac{1}{C_{\ell_1}^{TT,\mathrm{res}} + N_{\ell_1}^{TT}} \right) \nonumber \\ 
    &&\times \left(\frac{(C_{\ell_2}^{\phi\phi})^2 }{C_{\ell_2}^{\phi\phi}+N_{\ell_2}^{\phi\phi}}\right) \Gamma^{(x) \ell_1 \ell_2 1 *}_{m_1 m_2 m} \,, 
\ee
with the minimal reconstruction noise of  the dipole given by
\be
	&N_{\rm {Dipole}}^{\widehat{T} \widehat{T}}& =  \left[ \sum_{\ell_1 \ell_2} e_{\ell_1 \ell_2 1} \frac{(2\ell_1+1)(2\ell_2+1)}{4\pi} \left(J_{\ell_1 \ell_2 1}\right)^2 \right. \nonumber \\
	&&\times  \left. \left(\frac{1}{C_{\ell_1}^{TT,\mathrm{res}} + N_{\ell}^{TT}}\right) \left( \frac{(C_{\ell_2}^{\phi\phi})^2 }{C_{\ell_2}^{\phi\phi}+N_{\ell_2}^{\phi\phi}}\right)\right]^{-1}.\nonumber \\ 
    \label{eq:reconstructionnoise}
\ee 	
Here $C_{\ell_1}^{TT,\mathrm{res}}$ is the delensed temperature spectrum~\cite{Green:2016cjr}, the use of which lowers the variance of the reconstruction~\cite{Meerburg:2017lfh}. 

\section{Forecasts}
First, we  estimate the amount of information available in principle, i.e., given perfect data.  We assume we have measured the temperature and the lensing fields  to cosmic variance up to a given $\ell_{\rm max}$, and that we can delens perfectly. We show the  noise on the reconstructed dipole as a function of $\ell_{\rm max}$  in Fig.~\ref{fig:PerfectUniverseDipole}. Within  $\Lambda$CDM we predict a primordial dipole power of $C_1 = \langle |T_{1m}|^2 \rangle_m \sim 10^3$ $\mu$K$^2$.  Hence, we would have a $S/N \sim 1$ in this idealized configuration for  $\ell_{\rm max}\sim 5\times 10^4$ on both the temperature and the lensing field. The significant improvement that we observe between $\ell =3000$ and $\ell = 5000$ is due to the fact that on those scales lensing becomes increasingly important. 

In the same figure we show the  error on the kinematic dipole that can be achieved through a measurement of multipole aberration \cite{Kamionkowski:2002nd} with the methods developed in Ref.~\cite{Aghanim:2013suk}.  Future CMB experiments should be able to detect the kinematic dipole with high significance using this effect. A difference between the dipole measured through aberration and the measurement of the CMB dipole would be a probe of the primary dipole. We note that we only include the aberration effect from our motion and do not include the modulation effect, since as explained in Ref.~\cite{Aghanim:2013suk} the latter cannot be used to distinguish between a primary dipole and one from our motion in the local CMB rest frame.
In this idealized scenario, the reconstruction we propose compares favorably to the ability to constrain the intrinsic dipole using aberration effects, becoming more constraining at $\ell_\mathrm{max} \simeq 3000$ when combining all combinations of polarization and temperature in the aberration measurement.   We note that $EB$ and $TB$ do not contribute any significant weight to the reconstruction of the aberration effect (or lensing modes at $\ell=1$ in general)~\cite{Amendola:2010ty}.

\begin{figure}
\includegraphics[width=\columnwidth]{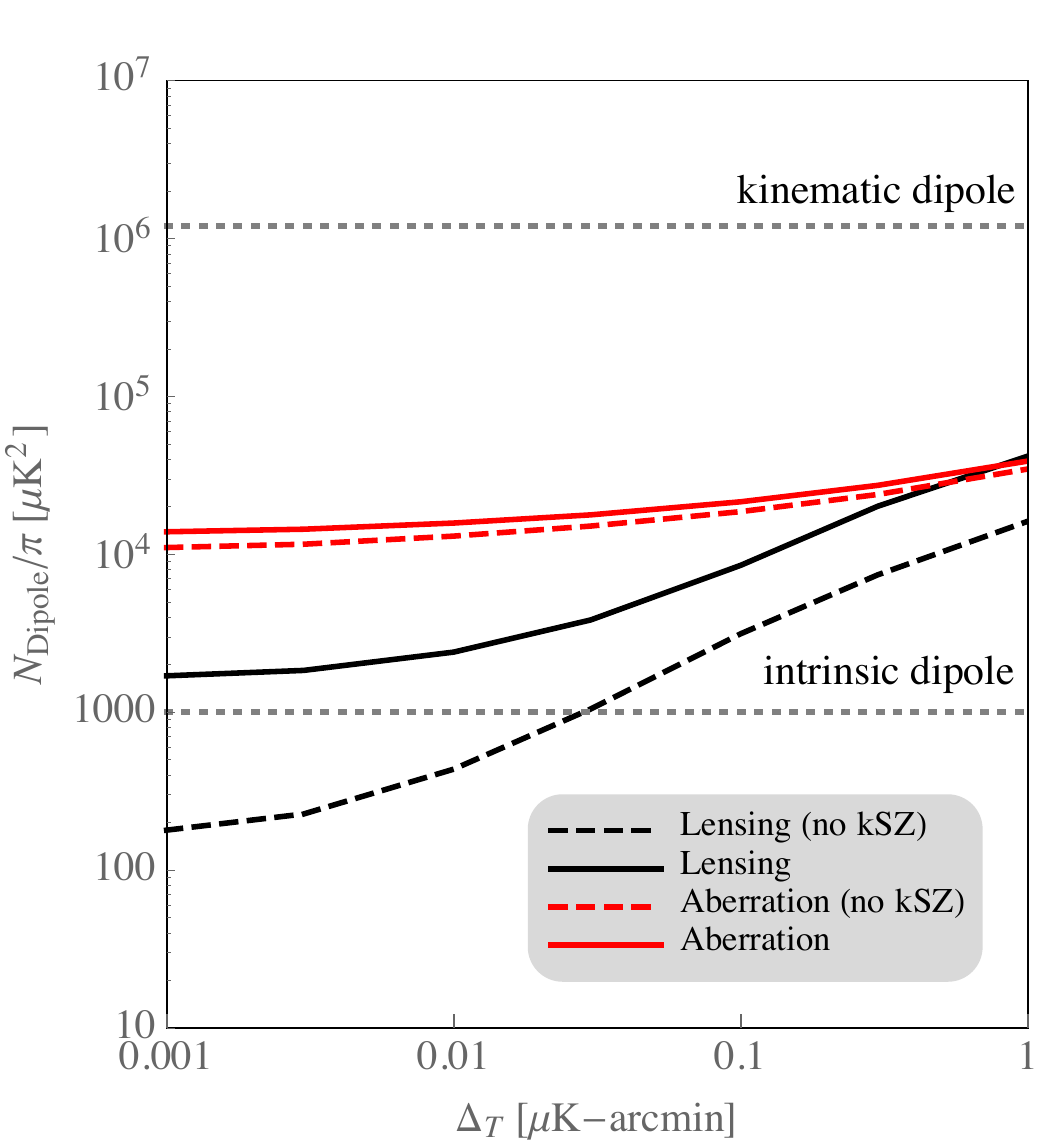}
\caption{
Noise on the reconstructed dipole as a function of experimental noise. In black we show the noise obtained lensing-based reconstruction method, using internal lensing reconstruction and delensing.  In red, we show the noise from the aberration-based kinematic reconstruction.  For both cases the the dashed line neglects kSZ power, while the solid line includes it.}
\label{fig:realisticDipoleNoise}
\end{figure}

Next, we  include realistic noise levels, lensing potential reconstruction, and delensing of the temperature maps on small scales.  Small-scale CMB maps contain fluctuations from galaxies and galaxy clusters.  Even if the frequency-dependent thermal Sunyaev-Zel'dovich effect and the emission from radio and star-forming galaxies are completely cleaned with multi-frequency observations, the kSZ effect will remain as it has the same frequency dependence as the primary CMB.  We thus include a model for the kSZ spectrum \cite{kSZpower2011}, which enters as effective noise in the temperature maps. In Fig.~\ref{fig:realisticDipoleNoise} we show the noise on the reconstructed dipole as a function of the noise level of a given experiment, where we assume that lensing reconstruction is performed internally at the given noise level. As found with polarization reconstruction in Ref.~\cite{Meerburg:2017lfh}, one main limitation is the lensing noise; the kSZ power spectrum  also significantly affects the results. All curves are derived with $\ell_{\rm min} = 20$ and a telescope beam with full-width at half maximum of 1$^\prime$.  The best constraint on the primary dipole that can be achieved with noise levels that are expected within the next decade, i.e. $\Delta_T \simeq 1\, \mu$K-arcmin \cite{CMBS4}, will come from residuals between the direct dipole and from  aberration effects. However, our reconstruction quickly improves when considering lower noise levels. Independent of the extent to which  the kSZ can be removed from the temperature map, we expect our method to outperform the aberration-based constraints in more futuristic CMB missions, and to reach cosmic variance limits on the primary dipole for a noise level of $\Delta_T \simeq 0.02\, \mu$K-arcmin if we could remove the kSZ realization entirely. 

\section{Application: Primordial Non-Gaussianity}
An active field in cosmology is the search for non-vanishing higher-ordered moments in the distribution of initial density fluctuations \cite{Bartolo:2010qu,Ade:2015ava,Alvarez:2014vva}.  
A common type of non-Gaussianity searched for is the local bispectrum, which is a measure of correlations between triplets of modes in which one of the fluctuations has much longer wavelength than the other two.
We now estimate the anticipated improvement for a measurement of the local bispectrum using a minimal multipole of $\ell_\mathrm{min} = 1$ rather than  $\ell_\mathrm{min} = 2$. 
The error on the non-Gaussianity amplitude for this shape, $f_{\rm NL}^\mathrm{local}$, is given by
\be
\sigma_{f_{\rm NL}^\mathrm{local}}  = \left[\sum_{\ell_{\rm min} \leq \ell_1 \leq \ell_2 \leq \ell_3 \leq \ell_{\rm max}} \frac{(B_{\ell_1 \ell_2 \ell_3})^2}{\sigma_{\ell_1\ell_2\ell_3 }^2}\right]^{-1/2},
\label{eq:SNbispectrum}
\ee
with $B_{\ell_1\ell_2\ell_3} = h_{\ell_1\ell_2\ell_3}b_{\ell_1 \ell_2 \ell_3}$, where 
\be
h_{\ell_1 \ell_2 \ell_3} &=&   \sqrt{\frac{(2\ell_1+1)(2\ell_2+1)(2\ell_3+1)}{4\pi}} \left( \begin{matrix} \ell_1 & \ell_2 & \ell_3 \\ 0 & 0 & 0 \end{matrix} \right),  \nonumber
\ee
and where the reduced local bispectrum $b_{\ell_1\ell_2\ell_3}$ is given by \cite{Fergusson:2008ra}
\be
b_{\ell_1\ell_2\ell_3}\propto \sum_{i\neq j\leq3} \frac{1}{\ell_i(\ell_i+1)\ell_j(\ell_j+1)}. 
\ee
The weighting on each triplet of multipoles is given by 
\be
\sigma^2_{\ell_1\ell_2\ell_3 } = C_{\ell_1}C_{\ell_2}C_{\ell_3}\Delta_{\ell_1\ell_2\ell_3},
\ee
with $\Delta_{\ell_1\ell_2\ell_3} = 1$, $2$, and 6 for three, two, and one unique values of $\ell$ respectively, and we approximate $C_{\ell}^{TT} \propto [\ell(\ell+1)]^{-1}$ as is valid on large scales. With these assumptions we use Eq.~\eqref{eq:SNbispectrum} for various values of $\ell_{
\rm min}$ and $\ell_{\rm max}$, summing over all triangle configurations and assuming minimal covariance with other shapes. We find that if the primary dipole is measured to cosmic variance limits, its contribution to the squeezed bispectrum can improve constraints on $f_{\rm NL}^{\rm local}$ by about 10$\%$.

\section{Discussion and Conclusions}
In this paper we investigated the possibility of reconstructing the intrinsic CMB dipole given a  map of the CMB on small scales together with a map of the lensing potential.  We first considered an ideal survey, without noise or foreground fluctuations; in this case the dipole can be reconstructed to cosmic variance limits if the CMB and lensing field are mapped to scales of $\ell_\mathrm{max} \sim 5000$, or about 2$^\prime$.  For finite noise, we found that the dipole could be reconstructed to cosmic variance limits for map noise levels of $0.02\; \mu$K-arcmin, depending on the degree to which foregrounds can be cleaned from the CMB.  This can be compared with the $1\; \mu$K-arcmin noise level for the upcoming CMB-S4 survey.  A survey that targets this signal would also need sufficient sky coverage.

We compared with an alternative approach for obtaining the primary dipole, based on the reconstructing our motion from the induced aberration and subtracting this from the direct measurement of the dipole.  In contrast with our lensing-based method, this aberration-based method does not reach cosmic variance limits for any of the scenarios we considered.  

An estimate of the dipole could help our effort to explore the physics of the early Universe. As a concrete example, we showed how constraints on squeezed non-Gaussianities  benefit from this single mode. Similarly, an anomalously large dipole could only be sourced by super-horizon isocurvature fluctuations, possibly providing strong constraints on multifield inflation.  

Our proposed method can be used to recover the intrinsic CMB dipole to cosmic variance limits for an ideal experiment, and thus provides a compelling target for futuristic CMB surveys.

\vspace{0.7em}

\noindent
{\bf Acknowledgments}
We would like to thank  Nick Battaglia,  Anthony Challinor, Jens Chluba, Simone Ferraro, Matt Johnson, Niels Oppermann, Guilherme Pimentel, Douglas Scott, Kendrick Smith, and David Spergel  for discussions.  J.M. was supported by the Vincent and Beatrice Tremaine Fellowship.

\appendix
\section{Appendix: Dipole Lensing}
In this appendix we provide some details about gravitational lensing of the CMB dipole.  We will show that lenses which are in the local CMB rest frame lens only the intrinsic dipole, while lenses moving with respect to the CMB cause lensing of both the intrinsic dipole and the kinematic dipole as seen from the rest frame of the lens, due to moving lens effects.

Let us begin by considering a universe in which the CMB is completely uniform when viewed from the CMB rest frame.  Now place a single lens, stationary in the CMB rest frame, somewhere in the universe.  Photons are deflected due to the presence of the lens, according to $\tilde{T}( \boldsymbol{\hat{n}} ) = T( \boldsymbol{\hat{n}} + \boldsymbol{\alpha} \left( \boldsymbol{\hat{n}} \right))$.
Since the CMB is uniform in this example, the CMB observed by an observer which is in the CMB rest frame (and thus stationary with respect to the lens also) remains unchanged, i.e. $\tilde{T} ( \boldsymbol{\hat{n}} ) = T( \boldsymbol{\hat{n}} ) = T_0$.  

Next consider a similar setup, except with an observer moving at a constant velocity $\boldsymbol{\beta}$ with respect to the CMB rest frame (and the lens). In this frame, the unlensed CMB temperature takes the form $T'(\boldsymbol{\hat{n}}') = \left( \frac{\nu'}{\nu} \right) T(\boldsymbol{\hat{n}})$. Here the frequencies are related
by~\cite{Challinor:2002zh,Weinberg:2008zzc,Amendola:2010ty}
\be 
	\nu'(\boldsymbol{\hat{n}}) = \nu \gamma \left( 1 + \boldsymbol{\hat{n}}\cdot \boldsymbol{\beta} \right) \, ,
\ee
where $\gamma = (1-\beta^2)^{-1/2}$, and the angles are related by
\be 
	\boldsymbol{\hat{n}}' = \frac{\boldsymbol{\hat{n}} \cdot \boldsymbol{\hat{\beta}} + \beta}{1+\boldsymbol{\hat{n}} \cdot \boldsymbol{\beta}} \boldsymbol{\hat{\beta}} + \frac{\boldsymbol{\hat{n}} - \left( \boldsymbol{\hat{n}} \cdot \boldsymbol{\hat{\beta}} \right) \boldsymbol{\hat{\beta}}} {\gamma \left( 1+\boldsymbol{\hat{n}} \cdot \boldsymbol{\beta} \right) } \, .
\ee
Since the lensed and unlensed maps are the same in the original frame, the Lorentz boost to the moving frame must maintain the equality $\tilde{T}' \left( \boldsymbol{\hat{n}}' \right) = T'\left( \boldsymbol{\hat{n}}' \right) = \left( \frac{\nu'}{\nu} \right) T_0$.  

\begin{figure}
\includegraphics[width=0.7\columnwidth]{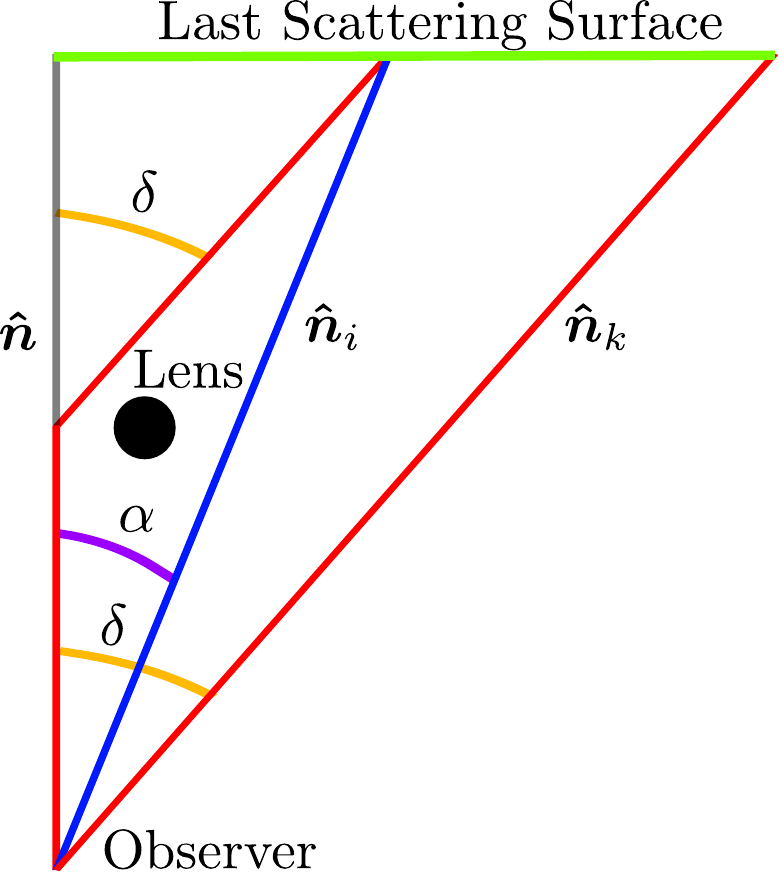}
\caption{Relevant geometry for dipole lensing.  For fluctuations intrinsic to the CMB, lensing redirects photons originating from the last scattering surface at a position $\boldsymbol{\hat{n}}_i$ in the unlensed map into the direction $\boldsymbol{\hat{n}}$. On the other hand, the kinematic dipole should be treated as a source at infinity, and so the temperature along $\boldsymbol{\hat{n}}_k$ in the unlensed map is observed along $\boldsymbol{\hat{n}}$ for lensing of the kinematic dipole.
}
\label{fig:LensingGeometry}
\end{figure}

It will be enlightening to confirm this result by calculating the effects of the moving lens directly in the frame of the  observer.  In this frame, the unlensed CMB temperature is given by $T\left( \boldsymbol{\hat{n}} \right) = \left( \frac{\nu'}{\nu} \right) T_0$, and the lens moves with velocity $\mathbf{v} = -\boldsymbol{\beta}$. 
The lensed temperature for small lensing deflections is given by~\cite{Lewis:2006fu}
\be  
 	\tilde{T}(\boldsymbol{\hat{n}}) &=&  T\left( \boldsymbol{\hat{n}} + \boldsymbol{\alpha}(\boldsymbol{\hat{n}}) \right) \nonumber \\
    &=& T \left( \boldsymbol{\hat{n}} \right)
    + \boldsymbol{\nabla} T( \boldsymbol{\hat{n}} ) \cdot \boldsymbol{\alpha}( \boldsymbol{\hat{n}} ) + \mathcal{O}(\alpha^2) \, .
\ee
In addition to the deflection, the fact that the lens is moving results in a frequency shift which can be interpreted as a time-dependence of the potential, similar to the Rees-Sciama and integrated Sachs-Wolfe effects~\cite{1983Natur.302..315B,1986Natur.324..349G,1989LNP...330...59B,Lewis:2006fu}. For small deflections, this frequency shift takes the form
\be 
	\frac{\Delta \nu(\boldsymbol{\hat{n}})}{\nu} = \gamma_v \mathbf{v}_\perp \cdot \mathbf{\delta}(\mathbf{\hat{n}})
\ee
where $\mathbf{v}$ 
is the lens velocity, $\gamma_v = (1-v^2)^{-1/2}$, with $\mathbf{v}_\perp$ the component perpendicular to the sky, and $\boldsymbol{\delta}$ is the deflection angle of the photon at the lens.  
 In the case at hand, the deflection and the frequency shift 
combine for a temperature perturbation of the form
\be 
	\Delta T (\boldsymbol{\hat{n}}) &=& T_0 \gamma \boldsymbol{\beta}_\perp \cdot \boldsymbol{\alpha}(\boldsymbol{\hat{n}}) + T_0 \gamma_v \mathbf{v}_\perp  \cdot \boldsymbol{\delta}(\boldsymbol{\hat{n}}) \nonumber \\
    &=& T_0 \gamma \boldsymbol{\beta}_\perp \cdot (\boldsymbol{\alpha}(\boldsymbol{\hat{n}}) - \boldsymbol{\delta}(\boldsymbol{\hat{n}})) \, ,
    \label{Eq:MovingLensCancel}
\ee
where in the second line we used the fact that the lens has velocity $\mathbf{v} = -\boldsymbol{\beta}$.  At first glance, these perturbations do not cancel in general, since $\boldsymbol{\alpha}$ is related to $\boldsymbol{\delta}$ by a ratio of source and lens distances.  However, the kinematic dipole is not an intrinsic property of the last scattering surface, and instead should be treated as a source at infinite distance~\cite{Cooray:2005my,Lewis:2006fu}.  This can be understood by noticing that the temperature due to the Doppler shift as viewed in a direction $\boldsymbol{\hat{n}} + \boldsymbol{\delta}(\boldsymbol{\hat{n}})$ from the position of the lens is the same as that viewed in a direction $\boldsymbol{\hat{n}} + \boldsymbol{\delta}(\boldsymbol{\hat{n}})$ from the position of the observer, and not in a direction $\boldsymbol{\hat{n}} + \boldsymbol{\alpha}(\boldsymbol{\hat{n}})$ as would be the case if one were considering intrinsic fluctuations at the last scattering surface (see Fig.~\ref{fig:LensingGeometry}) which is a necessary consequence of physical consistency.Once this identification is made, we find that the deflection cancels with the frequency shift such that the moving lens has no net effect on the temperature map $\Delta T (\boldsymbol{\hat{n}})=0$, giving a result consistent with what we found by Lorentz boosting the lensed temperature in the previous paragraph.

Let us summarize a few lessons that we learn from this calculation.  First, we see that our kinematic dipole is not lensed by gravitational potentials which are stationary in the CMB rest frame.  Incidentally, lenses which move with respect the CMB rest frame add temperature fluctuations on small scales, though the power from moving lenses is predicted to be smaller than that from the kSZ effect~\cite{Aghanim:1998ux}. Next, there is a physical distinction between the intrinsic dipole and kinematic dipole, as was argued above. Last, the intrinsic dipole is lensed, allowing for the reconstruction method described in the main text to be a useful probe of the intrinsic dipole.

\bibliography{bib}
\end{document}